# Shear viscosity of superfluid $^3He$-$A_1$ at low temperatures

M. A. Shahzamanian and R. Afzali

Department of Physics, Faculty of Sciences, University of Isfahan, 81744 Isfahan, IRAN
E-mail addresses: shahzamanian@hotmail.com
afzali2001@hotmail.com

Abstract

The shear viscosity tensor of the $A_1$ – phase of superfluid $^3He$ is calculated at low temperatures and melting pressure, by using Boltzmann equation approach. The two normal and superfluid components take part in elements of the shear viscosity tensor differently. The interaction between normal and Bogoliubov quasiparticles in the collision integrals is considered in the binary, decay and coalescence processes. We show that the elements of the shear viscosities $\eta_{xy}$, $\eta_{xz}$ and $\eta_{zz}$ are proportional to $(T/T_c)^{-2}$. The constant of proportionality is in nearly good agreement with the experimental results of Roobol et al.

## 1. Introduction

There are three distinct stable superfluid phases of bulk $^3He$, namely A, B and $A_1$ phases. In zero magnetic field, only the A and B phases are stable [1]. The $A_1$ phase appears as a narrow wedge between the normal state and the A and B phases. In the presence of a magnetic field, the A and B phases are referred to as $A_2$ and $B_2$ phases [1]. It is widely believed from the nuclear magnetic resonance experimental data and theoretical work that the A phase of liquid $^3He$ is the Anderson-Brinkman-Morel state [2, 3]. In this axial p-wave state only ($\uparrow\uparrow$) and ($\downarrow\downarrow$) pairs exist, and in weak coupling limit may be totally decoupled. In the presence of a static magnetic field the A phase of liquid $^3He$ splits in to two phases, $A_1$ and $A_2$, where in the $A_1$ phase only a single spin population is paired and the $A_2$ phase contains two independently paired spin populations [4]. Different theoretical researches have been so far conducted on the shear viscosity of a superfluid $^3He$, specially in zero magnetic field. The shear viscosity of isotropic superfluid state of $^3He$ has been discussed theoretically by Shumeiko [5]. Shumeiko, by taking the kinetic equation approach, showed that the shear viscosity of a superfluid s- wave state for temperatures close to the transition temperature $T_c$ behaves like $\eta = \eta_n(T_c) [1-c(\Delta/K_BT)]$ where $\eta_n(T_c)$ is the shear viscosity of the normal phase at the transition temperature $T_c$, $K_B$ is Baltzmann's constant, and c is a numerical constant. The shear viscosity of the A- phase, in zero magnetic field, has been calculated exactly for temperatures close to $T_c$ by Pethick et al [6], and Bhattacharyya et al [7] in s-p approximation. Their result on the viscosity drops as $(1-T/T_c)^{1/2}$ and the exact coefficient of $(1-T/T_c)^{1/2}$ has been expressed as a function of normal state properties. The shear viscosity of the B-phase, in zero magnetic field, at low temperatures, along with the calculation of transition probabilities in s-p approximation has already been investigated by Pethick et al [8]. A microscopic calculation of the fourth-rank shear viscosity tensors of superfluid phases of $^3He$, using Kubo formula approach, has been done by Shahzamanian [9]. The calculation of shear viscosity for $A_1$–phase, in nonzero magnetic field, has turned out to be more complicated than that for the other phases, due to the anisotropy of the gap. In high magnetic field, strong coupling effect must be considered for maximum gap parameter, and on normal fluid density components of the $A_1$ and $A_2$ phases of liquid $^3He$ [10]. The viscosity, in the presence of a magnetic field has been considered by Shahzamanian close to $T_c$ in relaxation time approximation [10]. Shahzamanian and Afzali [11] calculated the shear viscosity of $A_1$-phase close to $T_c$, considering superfluid-normal interaction along with other interactions for binary, decay and coalescence processes of quasiparticles, and we also moved beyond s-p approximation by the use of Pfitzner procedure. The shear viscosity of $^3He$-B has been measured down to temperatures $T/T_c \simeq 0.4$ at different pressures simultaneously by vibrating-wire and torsional-oscillator viscometers [12] and down to $T/T_c \simeq 0.3$ by a vibrating-wire viscometer [13]. Experimentally, the shear viscosity of the A phase has been measured by Alvesalo et al [14-16] in a vibrating string viscometer experiment and indirectly by Johnson et al [17]. The shear viscosity, in the presence of low magnetic field, has been measured by Alvesalo et al [16]. A number of viscosity measurements have been carried out in the $A_1$ and $A_2$ phases of superfluid $^3He$ at different high magnetic fields by Roobol et al [18-20]. All the results show a sharp decrease proportional to the opening of the superfluid energy gap, $\Delta \propto (1-T/T_c)^{1/2}$ near $T_{c_1}$ and goes as $T^{-2}$ for low temperatures.

In this paper, we study the shear viscosity of $A_1$-phase of superfluid $^3He$ at low temperatures, using Boltzmann equation approach. We assume that magnetic field is sufficiently high so that all of quasiparticles with spin up go to superfluid state and all quasiparticles with spin down stay in normal state.

In this paper also we consider normal-superfluid interactions which come from the scattering between superfluid quasiparticles in the spin-up population, the so-called Bogoliubov quasiparticles and the normal fluid quasiparticles in the spin-down population. In a normal Fermi liquid at low temperatures the only important collision process is the binary scattering of quasiparticles, but in a superfluid the quasiparticle number is not conserved and other processes as well as binary processes can occur. So we take into account these processes such as decay processes in which one Bogoliubov quasiparticle decays into three and coalescence processes in which three Bogoliubov quasiparticles coalesce to produce one.

Transition probabilities are explicitly extended in terms of spin indices. It has been cited elaborately how transition probabilities can be expressed in terms of singlet and triplet quasiparticle scattering amplitude (see section 2); and accordingly, Pfitzner procedure [21] has been used in the calculation of quasiparticle scattering amplitude. In Pfitzner procedure, necessary conditions are established to explicitly construct exchange-symmetric scattering amplitudes by adding higher angular momentum components. More clearly, Pfitzner introduces a general set of complete and orthonormalized eigenfunctions of the exchange operator on the Fermi sphere. Then, out of it, He manages to construct a general polynomial expansion of the quasiparticle scattering amplitude. By using this expansion, it is attempted to fit the quasiparticle scattering amplitude by including as many available experimental data (such as Landau parameters and transport times in the normal phase of $^3He$) as possible and taking account of their respective uncertainties. This expansion of quasiparticle scattering amplitude fully contains the s-p approximation as special case (see the sentences following Eq. (39) ) but it should be borne in mind that this expansion is absolutely general. Therefore, there maybe no need to say that s-p approximation is a simple approximation for quasiparticle scattering amplitude in terms of Landau parameters.

Another study is also under way dealing with intermediate temperatures regions, the complete solution of which requires numerical computations and will be published elsewhere.

The paper is organized as follows. In section 2, we obtain transition probabilities of different processes and discuss about those. In section 3, by writing collision integral, Boltzmann equation is solved for $A_1$-phase. In section 4, the shear viscosity has been calculated at low temperatures, and in section 5, we give some remarks and concluding results.

## 2. Transition probabilities

To obtain transition probabilities, we start with the interaction term in the Hamiltonian, i.e.

$$H = \frac{1}{4} \sum_{1,2,3,4} \langle 3,4|T|1,2 \rangle a^\dagger_4 a^\dagger_3 a_1 a_2 \tag{1}$$

where i = 1,2,3 and 4, stands for both momentum ($\vec{p}_i$) and spin ($\sigma_i$) variables.

Suppose that only spin up particles become superfluid. The creation and annihilation operators with spin down in $H$ are left intact. By using Bogoliubov transformation, the quasiparticle creation $a^\dagger_{\vec{p},\uparrow}$ and annihilation $a_{\vec{p},\uparrow}$ operators may be replaced by the creation and annihilation operators $\alpha^\dagger_{\vec{P},\sigma}$ and $\alpha_{\vec{P},\sigma}$ in the superfluid.

Bogoliubov transformation may be written as

$$a_{\vec{p},\sigma} = u_{\vec{p},\sigma\sigma'} \alpha_{\vec{p},\sigma'} - v_{\vec{p},\sigma\sigma'} \alpha^\dagger_{-\vec{p},\sigma'} \tag{2}$$

$$a^\dagger_{\vec{p},\sigma} = v^*_{\vec{p},\sigma\sigma'} \alpha_{\vec{p},\sigma'} + u_{\vec{p},\sigma\sigma'} \alpha^\dagger_{-\vec{p},\sigma'} \tag{3}$$

where the matrix elements $u_{\vec{p},\sigma\sigma'}$ and $v_{\vec{p},\sigma\sigma'}$ can be chosen for $A_1$-phase as [22]

$$u_{\vec{p},\sigma\sigma'} = \left[\frac{1}{2}\left(1+\varepsilon_{\vec{p}}/E_{\vec{p}}\right)\right]^{1/2}\delta_{\sigma\sigma'} \quad ; \quad v_{\vec{p},\sigma\sigma'} = \left[\frac{1}{2}\left(1-\varepsilon_{\vec{p}}/E_{\vec{p}}\right)\right]^{1/2}\delta_{\sigma\sigma'} \tag{4}$$

In the $A_1$-phase we may write $E_{\vec{p}} = \left(\varepsilon_{\vec{p}}^2 + \left|\Delta_{\vec{p}\uparrow\uparrow}\right|^2\right)^{1/2} \text{sgn}\,\varepsilon_{\vec{p}}$, where $\varepsilon_{\vec{p}}$ is the normal-state quasiparticle energy measured with respect to the chemical potential and $\Delta_{\vec{p}\uparrow\uparrow}$ is the magnitude of the gap in the direction $\vec{p}$ on the Fermi surface [22]. $\left|\Delta_{\vec{p}\uparrow\uparrow}\right|$ is equal to $\Delta(T)\sin\theta_p$, where $\Delta(T)$ is the maximum gap and $\theta_p$ is the angle between the quasiparticle momentum and gap axis $\hat{\ell}$ that suppose to be in the direction of z-axis. For the non-unitary state of the $A_1$-phase, there are the following properties between $u$ and $v$ [22].

$$v_{-\vec{p},\uparrow\uparrow} = -v_{\vec{p},\uparrow\uparrow} \quad ; \quad u_{-\vec{p},\uparrow\uparrow} = u_{\vec{p},\uparrow\uparrow} \tag{5}$$

Then, by using Eqs.(2) and (3) in Eq.(1) we have

$$H = \frac{1}{4}\sum_{\vec{P}_1,\vec{P}_2,\vec{P}_3,\vec{P}_4}\left\{\left[\langle 3\uparrow 4\uparrow|T|1\uparrow 2\uparrow\rangle\left(u_{4\uparrow\uparrow}\alpha_{4\uparrow}^\dagger - v_{4\uparrow\uparrow}^*\alpha_{-4\uparrow}\right)\left(u_{3\uparrow\uparrow}\alpha_{3\uparrow}^\dagger - v_{3\uparrow\uparrow}^*\alpha_{-3\uparrow}\right)\left(u_{1\uparrow\uparrow}\alpha_{1\uparrow} - v_{1\uparrow\uparrow}\alpha_{-1\uparrow}^\dagger\right)\left(u_{2\uparrow\uparrow}\alpha_{2\uparrow} - v_{2\uparrow\uparrow}\alpha_{-2\uparrow}^\dagger\right)\right]$$

$$+\left[\langle 3\downarrow 4\downarrow|T|1\downarrow 2\downarrow\rangle a_{4\downarrow}^\dagger a_{3\downarrow}^\dagger a_{1\downarrow} a_{2\downarrow}\right]$$

$$+\left[\langle 3\downarrow 4\uparrow|T|1\uparrow 2\downarrow\rangle\left(u_{4\uparrow\uparrow}\alpha_{4\uparrow}^\dagger - v_{4\uparrow\uparrow}^*\alpha_{-4\uparrow}\right)a_{3\downarrow}^\dagger\left(u_{1\uparrow\uparrow}\alpha_{1\uparrow} - v_{1\uparrow\uparrow}\alpha_{-1\uparrow}^\dagger\right)a_{2\downarrow}\right]$$

$$+\left[\langle 3\uparrow 4\downarrow|T|1\uparrow 2\downarrow\rangle a_{4\downarrow}^\dagger\left(u_{3\uparrow\uparrow}\alpha_{3\uparrow}^\dagger - v_{3\uparrow\uparrow}^*\alpha_{-3\uparrow}\right)\left(u_{1\uparrow\uparrow}\alpha_{1\uparrow} - v_{1\uparrow\uparrow}\alpha_{-1\uparrow}^\dagger\right)a_{2\downarrow}\right]$$

$$+\left[\langle 3\downarrow 4\uparrow|T|1\downarrow 2\uparrow\rangle\left(u_{4\uparrow\uparrow}\alpha_{4\uparrow}^\dagger - v_{4\uparrow\uparrow}^*\alpha_{-4\uparrow}\right)a_{3\downarrow}^\dagger a_{1\downarrow}\left(u_{2\uparrow\uparrow}\alpha_{2\uparrow} - v_{2\uparrow\uparrow}\alpha_{-2\uparrow}^\dagger\right)\right]$$

$$+\left[\langle 3\uparrow 4\downarrow|T|1\downarrow 2\uparrow\rangle a_{4\downarrow}^\dagger\left(u_{3\uparrow\uparrow}\alpha_{3\uparrow}^\dagger - v_{3\uparrow\uparrow}^*\alpha_{-3\uparrow}\right)a_{1\downarrow}\left(u_{2\uparrow\uparrow}\alpha_{2\uparrow} - v_{2\uparrow\uparrow}\alpha_{-2\uparrow}^\dagger\right)\right]\right\} \tag{6}$$

which contains terms $\alpha_4^\dagger\alpha_3^\dagger\alpha_{-2}^\dagger\alpha_1$, $\alpha_4^\dagger\alpha_3^\dagger\alpha_2\alpha_1$, $\alpha_4^\dagger\alpha_1\alpha_{-3}\alpha_2$, $\alpha_4^\dagger\alpha_3^\dagger\alpha_{-2}^\dagger\alpha_{-1}^\dagger$ and $\alpha_{-4}\alpha_{-3}\alpha_2\alpha_1$. These terms decay one quasiparticle into three, convert two quasiparticles into two, coalescence three quasiparticles into one, create four quasiparticles from the condensate and scatter four quasiparticles into the condensate, respectively. The last two processes are not allowed, because in each process the total energy should be conserved.

The first bracket of Hamiltonian in Eq. (6) hence is related to the superfluid component interactions. The second bracket is related to normal component interactions which is important and included in this study. The other terms are related to normal-superfluid interaction and vice versa.

The transition probabilities for the processes may be defined, for example, as $|\langle \vec{p}_4,\uparrow;\vec{p}_3,\uparrow|H|\vec{p}_1,\uparrow;\vec{p}_2,\uparrow\rangle|^2$. By using Wick's theorem, we obtain the following transition probabilities relating to the T-matrix (throughout of this paper we put $\hbar \equiv K_B \equiv 1$). For normal component interaction, we have

$$W_{22}(\downarrow\downarrow) = 2\pi \ |\langle 4\downarrow 3\downarrow|T|1\downarrow 2\downarrow\rangle|^2 \equiv 2\pi |T_{t_I}|^2 \tag{7}$$

where subscripts on W indicate binary processes in which two quasiparticles with spin down scatter to two quasiparticles with spin down and $T_{t_I}$ is triplet quasiparticle scattering amplitude. $T_{t_I}$ is calculated by Pfitzner procedure [21] (see Eq. (38)). For superfluid component interaction, by using relation (5), we have

$$W_{22}(\uparrow\uparrow) = 2\pi \Big\{ \Big[ \big(|v_1|^2|v_2|^2|v_3|^2|v_4|^2 + |u_1|^2|u_2|^2|u_3|^2|u_4|^2 + 2u_1v_1u_2v_2u_3v_3u_4v_4\big)|\langle 4\uparrow 3\uparrow|T|1\uparrow 2\uparrow\rangle|^2 \ \Big]$$

$$-\Big[ \big(|v_1|^2|v_4|^2 u_2v_2u_3v_3 + |v_2|^2|v_3|^2 u_1v_1u_4v_4 + |u_2|^2|u_3|^2 u_1v_1u_4v_4 + |u_1|^2|u_4|^2 u_2v_2u_3v_3\big)$$

$$\times \big(\langle 4\uparrow 3\uparrow|T|1\uparrow 2\uparrow\rangle^*\langle -1\uparrow 3\uparrow|T|-4\uparrow 2\uparrow\rangle + \langle 4\uparrow 3\uparrow|T|1\uparrow 2\uparrow\rangle\langle -1\uparrow 3\uparrow|T|-4\uparrow 2\uparrow\rangle^*\big) \ \Big]$$

$$+\Big[ \big(|v_4|^2|v_2|^2 u_1v_1u_3v_3 + |v_1|^2|v_3|^2 u_2v_2u_4v_4 + |u_1|^2|u_3|^2 u_2v_2u_4v_4 + |u_2|^2|u_4|^2 u_1v_1u_3v_3\big)$$

$$\times \big(\langle 4\uparrow 3\uparrow|T|1\uparrow 2\uparrow\rangle^*\langle -2\uparrow 3\uparrow|T|-4\uparrow 1\uparrow\rangle + \langle 4\uparrow 3\uparrow|T|1\uparrow 2\uparrow\rangle\langle -2\uparrow 3\uparrow|T|-4\uparrow 1\uparrow\rangle^*\big) \ \Big]$$

$$+ \Big[ \big(|v_1|^2|u_3|^2|u_2|^2|v_4|^2 + |v_2|^2|u_4|^2|u_1|^2|v_3|^2 + 2u_1v_1u_2v_2u_3v_3u_4v_4\big)|\langle -1\uparrow 3\uparrow|T|-4\uparrow 2\uparrow\rangle|^2 \ \Big]$$

$$-\Big[ \big(|u_3|^2|v_4|^2 u_1v_1u_2v_2 + |v_1|^2|u_2|^2 u_3v_3u_4v_4 + |u_1|^2|v_2|^2 u_3v_3u_4v_4 + |v_3|^2|u_4|^2 u_1v_1u_2v_2\big)$$

$$\times \big(\langle -1\uparrow 3\uparrow|T|-4\uparrow 2\uparrow\rangle^*\langle -2\uparrow 3\uparrow|T|-4\uparrow 1\uparrow\rangle + \langle -1\uparrow 3|T|-4\uparrow 2\uparrow\rangle\langle -2\uparrow 3\uparrow|T|-4\uparrow 1\uparrow\rangle^*\big) \ \Big]$$

$$+\Big[ \big(|u_1|^2|v_2|^2|u_3|^2|v_4|^2 + |v_1|^2|u_2|^2|v_3|^2|u_4|^2 + 2u_1v_1u_2v_2u_3v_3u_4v_4\big)|\langle -2\uparrow 3\uparrow|T|-4\uparrow 1\uparrow\rangle|^2 \ \Big] \Big\}$$

(8)

The transition probability for decay process may be written as

$$W_{13}(\uparrow\uparrow) = |\langle \vec{p}_3\uparrow,\vec{p}_4\uparrow,-\vec{p}_2\uparrow|H|\vec{p}_1\uparrow\rangle|^2$$

or

$$W_{13}(\uparrow\uparrow) = 2\pi \ |[(v_1^*v_2u_3v_4 - u_1u_2v_3u_4)\langle -1\uparrow 3\uparrow|T|-4\uparrow 2\uparrow\rangle$$
$$+ (v_1^*u_2v_3v_4 - u_1v_2u_3u_4)\langle 4\uparrow 3\uparrow|T|1\uparrow 2\uparrow\rangle$$
$$+ (v_1^*v_2v_3u_4 - u_1u_2u_3v_4)\langle 4\uparrow -1\uparrow|T|-3\uparrow 2\uparrow\rangle]|^2 \tag{9}$$

Similarly the transition probability for the coalescence process may be written as

$$W_{31}(\uparrow\uparrow) = |\langle \vec{p}_4 \uparrow |H| \vec{p}_1 \uparrow, \vec{p}_2 \uparrow, -\vec{p}_3 \uparrow\rangle|^2$$

or

$$W_{31}(\uparrow\uparrow) = 2\pi |[(v_1^* v_2^* u_3 v_4 - u_1 u_2 v_3^* u_4)\langle 4\uparrow 3\uparrow |T| 1\uparrow 2\uparrow\rangle$$
$$+ (v_1^* u_2 v_3^* v_4 - u_1 v_2^* u_3 u_4)\langle -1\uparrow 3\uparrow |T| -4\uparrow 2\uparrow\rangle$$
$$+ (u_1 v_2^* v_3^* v_4 - v_1^* u_2 u_3 u_4)\langle 4\uparrow -1\uparrow |T| -3\uparrow 2\uparrow\rangle]|^2 \tag{10}$$

For the superfluid-normal component interaction one has

$$W_{22}(\uparrow\downarrow) = 2\pi\{|u_1|^2 |u_3|^2 |\langle 4\downarrow 3\uparrow |T| 1\uparrow 2\downarrow\rangle|^2 + |v_1|^2 |v_3|^2 |\langle 4\downarrow -1\uparrow |T| -3\uparrow 2\downarrow\rangle|^2$$
$$- u_1^* v_1^* u_3 v_3 \langle 4\downarrow 3\uparrow |T| 1\uparrow 2\downarrow\rangle^* \langle 4\downarrow -1\uparrow |T| -3\uparrow 2\downarrow\rangle$$
$$- u_1 v_1 u_3^* v_3^* \langle 4\downarrow 3\uparrow |T| 1\uparrow 2\downarrow\rangle \langle 4\downarrow -1\uparrow |T| -3\uparrow 2\downarrow\rangle^*$$
$$+ |u_1|^2 |u_4|^2 |\langle 4\uparrow 3\downarrow |T| 1\uparrow 2\downarrow\rangle|^2 + |v_1|^2 |v_4|^2 |\langle -1\uparrow 3\downarrow |T| -4\uparrow 2\downarrow\rangle|^2$$
$$- u_1^* v_1^* u_4 v_4 \langle 4\uparrow 3\downarrow |T| 1\uparrow 2\downarrow\rangle^* \langle -1\uparrow 3\downarrow |T| -4\uparrow 2\downarrow\rangle$$
$$- u_1 v_1 u_4^* v_4^* \langle 4\uparrow 3\downarrow |T| 1\uparrow 2\downarrow\rangle \langle -1\uparrow 3\downarrow |T| -4\uparrow 2\downarrow\rangle^*\} \tag{11}$$

$W_{22}(\downarrow\uparrow)$ is obtained by replacing 2 with 1 and vice versa in Eq.(11).

It is noted that decay of quasiparticle with spin up is impossible because in such a condition, two outgoing quasiparticles out of three must have spin down. This shows the creation of quasiparticles of normal state which is not allowed, since the number of quasiparticles in normal state is conserved. Then we have the following result for decay of a quasiparticle with spin down.

$$W_{13}(\downarrow\uparrow) = 2\pi |\langle \vec{p}_3 \uparrow, \vec{p}_4 \uparrow, -\vec{p}_2 \downarrow |H| \vec{p}_1 \downarrow\rangle|^2$$
$$= 2\pi\{|u_3|^2 |v_4|^2 |\langle -2\downarrow 3\uparrow |T| -4\uparrow 1\downarrow\rangle|^2 + |v_3|^2 |u_4|^2 |\langle -2\downarrow 4\uparrow |T| -3\uparrow 1\downarrow\rangle|^2$$
$$+ u_3 v_3 u_4^* v_4^* \langle -2\downarrow 3\uparrow |T| -4\uparrow 1\downarrow\rangle^* \langle 4\uparrow -2\downarrow |T| -3\uparrow 1\downarrow\rangle$$
$$+ u_3^* v_3^* u_4 v_4 \langle -2\downarrow 3\uparrow |T| -4\uparrow 1\downarrow\rangle \langle 4\uparrow -2\downarrow |T| -3\uparrow 1\downarrow\rangle^*\} \tag{12}$$

The final quasiparticle in the coalescence process of the superfluid-normal interactions also must have spin down. Hence, we have

$$W_{31}(\downarrow\uparrow) = 2\pi \left|\langle \vec{p}_1 \downarrow |H| -\vec{p}_2 \downarrow, \vec{p}_3 \uparrow, \vec{p}_4 \uparrow \rangle\right|^2$$

$$= 2\pi \{|v_3|^2 |u_4|^2 \left|\langle 1\downarrow -3\uparrow |T| 4\uparrow -2\downarrow\rangle\right|^2 + |v_4|^2 |u_3|^2 \left|\langle -4\uparrow 1\downarrow |T| 3\uparrow -2\downarrow\rangle\right|^2$$

$$+ v_3 u_3 u_4^* v_4^* \langle 1\downarrow -3\uparrow |T| 4\uparrow -2\downarrow\rangle^* \langle -4\uparrow 1\downarrow |T| 3\uparrow -2\downarrow\rangle$$

$$+ v_3^* u_3^* u_4 v_4 \langle 1\downarrow -3\uparrow |T| 4\uparrow -2\downarrow\rangle \langle -4\uparrow 1\downarrow |T| 3\uparrow -2\downarrow\rangle^* \} \tag{13}$$

T- matrix elements which appear in the transition probabilities may be expressed in terms of the scattering amplitudes for pairs of quasiparticles in singlet and triplet states, $T_s$ and $T_t$ respectively. By considering properties of $T_s$ and $T_t$ [23], we have

$$\langle 4\downarrow 3\downarrow |T| 1\downarrow 2\downarrow\rangle = \langle 4\uparrow 3\uparrow |T| 1\uparrow 2\uparrow\rangle \equiv T_{t_I}$$

$$\langle 4\uparrow -1\uparrow |T| -3\uparrow 2\uparrow\rangle \equiv T_{t_{II}} \quad , \quad \langle -1\uparrow 3\uparrow |T| -4\uparrow 2\uparrow\rangle \equiv T_{t_{III}}$$

$$\langle 4\downarrow 3\uparrow |T| 1\uparrow 2\downarrow\rangle = \langle 4\uparrow 3\downarrow |T| 1\downarrow 2\uparrow\rangle \equiv \frac{1}{2}(-T_{s_I} + T_{t_I})$$

$$\langle 4\uparrow 3\downarrow |T| 1\uparrow 2\downarrow\rangle = \langle 4\downarrow 3\uparrow |T| 1\downarrow 2\uparrow\rangle \equiv \frac{1}{2}(T_{s_I} + T_{t_I})$$

$$\langle 4\uparrow -1\uparrow |T| -3\uparrow 2\downarrow\rangle = \langle 4\uparrow -1\downarrow |T| -3\downarrow 2\uparrow\rangle \equiv \frac{1}{2}(-T_{s_{II}} + T_{t_{II}})$$

$$\langle 4\uparrow -1\downarrow |T| -3\uparrow 2\downarrow\rangle = \langle 4\downarrow -1\uparrow |T| -3\downarrow 2\uparrow\rangle \equiv \frac{1}{2}(T_{s_{II}} + T_{t_{II}})$$

$$\langle -1\downarrow 3\uparrow |T| -4\uparrow 2\downarrow\rangle = \langle -1\uparrow 3\downarrow |T| -4\downarrow 2\uparrow\rangle \equiv \frac{1}{2}(-T_{s_{III}} + T_{t_{III}})$$

$$\langle -1\uparrow 3\downarrow |T| -4\uparrow 2\downarrow\rangle = \langle -1\downarrow 3\uparrow |T| -4\downarrow 2\uparrow\rangle \equiv \frac{1}{2}(T_{s_{III}} + T_{t_{III}}) \tag{14}$$

where $T_{t_I}$ and $T_{s_I}$ are given by [21]

$$N(0) T_{t_I, s_I}(\nu, P) = \sum_{k=0}^{\infty} \sum_{l=0}^{k} a_{lk} X_{lk}(\nu, P) \qquad (l \text{ even, odd}) \tag{15}$$

and $N(0) = m^* p_F / \pi^2$ is the density of states at the Fermi level. The coefficients with $l$ even (odd) belong to the singlet (triplet) part of the quasiparticle scattering amplitude. $X_{lk}$ appearing in Eq.(15) is

$$X_{lk}(\nu, P) = (k+1)^{1/2} (2l+1)^{1/2} (P^2/4 - 1)^l P_l(\nu) P_{k-l}^{(2l+1,0)}(P^2/2 - 1) \tag{16}$$

$$k = 0,1,..., \quad ; \quad l = 0,1,...,k$$

where the $P_l(\nu)$ and $P_n^{(a,b)}(x)$ are the Legendre polynomials and the Jacobi polynomials respectively, and finally

$$P \equiv 2\cos\frac{\theta}{2} \quad , \quad \nu \equiv \cos\varphi \tag{17}$$

where $\theta$ is the angle between the momenta of the incoming particles namely $\vec{p}_1$ and $\vec{p}_2$, and $\varphi$ is the angle between the planes spanned by the momentum vectors of the incoming particles and the outgoing particles. Similar equations for $T_{t_{II}}$ and $T_{s_{II}}$ ( $T_{t_{III}}$ and $T_{s_{III}}$ ) with replacement of $\cos\theta$ by $\cos\theta_{II}$ ($\cos\theta_{III}$) and $\cos\varphi$ by $\cos\varphi_{II}$ ($\cos\varphi_{III}$) can be used.

The following equations express $\theta_{II}, \varphi_{II}, \theta_{III}$ and $\varphi_{III}$ in terms of $\theta$ and $\varphi$ [8].

$$\cos\theta_{II} = -\cos^2\frac{\theta}{2} + (1-\cos^2\frac{\theta}{2})\cos\varphi \quad ; \quad \cos\theta_{III} = -\cos^2\frac{\theta}{2} - (1-\cos^2\frac{\theta}{2})\cos\varphi$$

$$\cos\varphi_{II} = \frac{(1-\cos^2\frac{\theta}{2})\cos\varphi + 3\cos^2\frac{\theta}{2} - 1}{(-1+\cos^2\frac{\theta}{2})\cos\varphi + 1 + \cos^2\frac{\theta}{2}} \quad ; \quad \cos\varphi_{III} = \frac{(1-\cos^2\frac{\theta}{2})\cos\varphi - 3\cos^2\frac{\theta}{2} + 1}{(1-\cos^2\frac{\theta}{2})\cos\varphi + 1 + \cos^2\frac{\theta}{2}}$$

$$\tag{18}$$

At low temperatures, we have $\sin\theta_{p_i} \simeq 0$ ($i=1,2,3,4$) [24], then we may write $\nu \simeq 0$ and $u \simeq 1$. It is noted that if we apply other approximations for $\nu$ and $u$, the temperature dependence does not change, however some small changes appear in the numerical coefficients of the transition probabilities. By substituting the values of $\nu$ and $u$ in the equations of the transition probabilities' (Eqs.(7)-(13)), we obtain

$$W_{22}(\downarrow\downarrow) = 2\pi |T_{t_I}|^2 \tag{19}$$

$$W_{22}(\uparrow\uparrow) \simeq 2\pi |T_{t_I}|^2 + \frac{\pi}{4}\left(|T_{t_{II}}|^2 + |T_{t_{III}}|^2\right) \tag{20}$$

$$W_{22}(\uparrow\downarrow) = W_{22}(\downarrow\uparrow) \simeq \frac{\pi}{2}\left(|-T_{s_I} + T_{t_I}|^2 + |T_{s_I} + T_{t_I}|^2\right) \tag{21}$$

and the other transition probabilities have nearly zero value. It is noted that only binary processes are dominated at low temperatures, and this is also the case for calculating the thermal diffusion coefficient of the A-phase at low temperatures [24].

3. Collision integrals.

The kinetic equation for the quasiparticle distribution function of superfluid, $\nu_{p,\sigma}$, may be written as [1]

$$\frac{\partial \nu_{p,\sigma}}{\partial t} + \frac{\partial \nu_{p,\sigma}}{\partial \vec{r}} \cdot \frac{\partial E_{p,\sigma}}{\partial \vec{p}} - \frac{\partial \nu_{p,\sigma}}{\partial \vec{p}} \cdot \frac{\partial E_{p,\sigma}}{\partial \vec{r}} = I(\nu_{p,\sigma}) \tag{22}$$

where $I(\nu_{p,\sigma})$ is the collision integral. Let us assume a disturbance of the form

$$\nu_\sigma(p,r) = \nu^0(p) + \delta\nu_\sigma(p) \tag{23}$$

where, in general, both $v^0(p)$ and $\delta v_\sigma(p)$ are functions of the variable r. We define the function $\psi_\sigma(p)$ by

$$\delta v_\sigma(p) = -\frac{1}{T} v^0(p)(1-v^0(p))\psi_\sigma(p) \tag{24}$$

where $v^0(p) = \left[\exp(E_p^0 - \vec{p}\cdot\vec{u})/T + 1\right]^{-1}$, $u$ is the slightly inhomogeneous velocity in the liquid. By substituting (23) in (22), keeping the terms which contribute to the shear viscosity, and supposing $\vec{u}$ and $\vec{\nabla}\cdot\vec{u}$ are zero at the point considered, to first order in $\delta v_\sigma(p)$ we have for initial quasiparticle with spin up [1]

$$-\frac{1}{2}\frac{\partial v^0}{\partial E_{p,\uparrow}} p_i \frac{\partial E_{p,\uparrow}}{\partial p_k}\left(\frac{\partial u_i}{\partial r_k} + \frac{\partial u_k}{\partial r_i} - \frac{2}{3}\delta_{ij}\vec{\nabla}\cdot\vec{u}\right) = I(\delta v_\uparrow(p)) \tag{25}$$

where $I(\delta v_\uparrow(p))$ is nearly equal to $I_{22}(\uparrow\downarrow)$. Eq. (25) for initial quasiparticle with spin down is given by changing the distribution function of superfluid state, $v^0(p)$, to the distribution function of normal state, $n^0(p)$, and $I(\delta v_\uparrow(p))$ to $I(\delta v_\downarrow(p))$. $I(\delta v_\downarrow(p))$ is equal to $I_{22}(\downarrow\downarrow) + I_{22}(\downarrow\uparrow)$.

Now by using (23) and keeping the terms to first order in $\psi_\sigma(p)$ and using Eqs. (19)-(21), the linearised collision terms in the Boltzmann equation may be written as:

$$I_{22}(\uparrow\downarrow) = \frac{(m^*)^3 T}{64\pi^5}\int \frac{\sin\theta}{\cos\frac{\theta}{2}} d\theta d\varphi d\varphi_2 \int_{-\infty}^{+\infty} dxdy\,(\psi(t) + \psi(x+y-t) - \psi(x) - \psi(y))$$

$$\times\left[\left|-T_{S_I} + T_{t_I}\right|^2 v^0(t)n^0(x+y-t)(1-v^0(x))(1-n^0(y))\right.$$

$$\left. + \left|T_{S_I} + T_{t_I}\right|^2 v^0(t)n^0(x+y-t)(1-n^0(x))(1-v^0(y))\right] \tag{26}$$

$$I_{22}(\downarrow\downarrow) = \frac{(m^*)^3 T}{16\pi^5}\int \frac{\sin\theta}{\cos\frac{\theta}{2}}|T_{t_I}|^2 d\theta d\varphi d\varphi_2 \int_{-\infty}^{+\infty} dxdy$$

$$\times n^0(t)n^0(x+y-t)(1-n^0(x))(1-n^0(y))(\psi(t) + \psi(x+y-t) - \psi(x) - \psi(y)) \tag{27}$$

where $\varphi_2$ is azimuthal coordinate of $\vec{p}_2$ relative to $\vec{p}_1$, $t \equiv E_1/T$, $x \equiv E_3/T$ and $y \equiv E_4/T$. To solve the linearised Boltzmann equation, it is suitable to define $q(t)$ as

$$\psi(t) = p_i \frac{\partial E_1}{\partial p_{1k}} q(t)\left[\frac{\partial u_i}{\partial r_k} + \frac{\partial u_k}{\partial r_i} - \frac{2}{3}\delta_{ij}\vec{\nabla}\cdot\vec{u}\right] \tag{28}$$

After expressing the bracket in Eq. (28) in terms of a series of spherical harmonics, i.e. $\sum_{m=-2}^{2} U_m p_2^{|m|}(\cos\Theta)e^{im\Phi}$ [25] and substituting (28) in the collision integrals, then integration on $\varphi_2$ can be done conveniently. By noting some symmetries with respect to variables in the collision integrals [25] and doing integration on y, then we get

$$I_{22}(\uparrow\downarrow) = \frac{(m^*)^3 T^2}{32\pi^4} \sum_{m=-2}^{2} U_m p_2^{|m|}(\cos\Theta)e^{im\Phi}\left(\frac{-\partial v^0}{\partial E_1}\right)\int dx K(t,x)\int \frac{\sin\theta d\theta d\varphi}{\cos\frac{\theta}{2}}$$

$$\times \left(|-T_{s_t}+T_{t_t}|^2 + |T_{s_t}+T_{t_t}|^2\right)\left[q(t)+q(-x)p_2(\cos\theta)-q(x)\{p_2(\cos\theta_{13})+p_2(\cos\theta_{14})\}\right]$$
(29)

where $\theta_{1i}$ is the angle between $\vec{p}_1$ and $\vec{p}_i$, and $\theta_{13}$ and $\theta_{14}$ are related to $\theta$ and $\varphi$ with the following relations.

$$\cos\theta_{13} = \cos^2\frac{\theta}{2} + \sin^2\frac{\theta}{2}\cos\varphi \quad ; \quad \cos\theta_{14} = \cos^2\frac{\theta}{2} - \sin^2\frac{\theta}{2}\cos\varphi \tag{30}$$

and $K(t,x)$ is

$$K(t,x) = \frac{e^{-t}+1}{e^{-x}+1} \frac{x-t}{e^{(x-t)}-1} \tag{31}$$

By substituting Eq. (29) in Eq. (25) and considering $K(t,x) = K(-t,-x)$, we have [25]

$$\int dx K(t,x)\left\{q_{s_\sigma}(t) - \lambda_{2s_\sigma} q_{s_\sigma}(x)\right\} = B_\sigma \tag{32}$$

$$\int dx K(t,x)\left\{q_{a_\sigma}(t) - \lambda_{2s_\sigma} q_{a_\sigma}(x)\right\} = O(TB_\sigma) \tag{33}$$

where $q_{s_\sigma}(x)$ and $q_{a_\sigma}(t)$ are symmetric and antisymmetric part of $q(t)$ respectively and subscript $\sigma$ may be indicated by $\uparrow$ or $\downarrow$. At low temperatures, Eq.(32) is dominated and one can ignore Eq. (33) with respect to it [25]. In Eq. (32), $\lambda_{2s_\sigma}$ and $B_\sigma$ are

$$\lambda_{2s_\sigma} = \int \frac{\sin\theta d\theta d\varphi}{\cos\frac{\theta}{2}} W_{22}\left\{1-\frac{3}{4}(1-\cos\theta)^2\sin^2\varphi\right\} \Bigg/ \int \frac{\sin\theta d\theta d\varphi}{\cos\frac{\theta}{2}} W_{22} \tag{34}$$

and

$$B_\sigma = \frac{16\pi^5}{m^{*3}T^2}\left[\int \frac{\sin\theta d\theta d\varphi}{\cos\frac{\theta}{2}} W_{22}\right]^{-1} \tag{35}$$

where for $\sigma = \uparrow$ and $\downarrow$ the function $W_{22}$ stands for $\frac{1}{2}W_{22}(\uparrow\downarrow)$ and $\frac{1}{4}W_{22}(\downarrow\downarrow)+\frac{1}{2}W_{22}(\downarrow\uparrow)$ respectively.

For obtaining $\lambda_{2s_\sigma}$ and $B_\sigma$, one must indicate the functional relations of $T_{t_I}, T_{s_I}, T_{t_{II}}$ and $T_{t_{III}}$ on $\theta$ and $\varphi$. Furthermore we note that $\theta$ is small for superfluid case and its maximum value is $\pi T/\Delta(0)$ [24] where maximum gap, $\Delta(0)$, due to strong coupling effects is equal to $1.77T_c$ [26]. Then we have

$$N(0)T_{s_I} = 2.47 + 6.61\left(\cos^2\frac{\theta}{2}\right) + 17.69\left(\cos^4\frac{\theta}{2}\right) - 11.2\left(\cos^6\frac{\theta}{2}\right)$$

$$+ (3\cos^2\varphi - 1)\sin^4\frac{\theta}{2}\left[2.9 - 0.96\left(7\cos^2\frac{\theta}{2} - 1\right)\right] \quad (36)$$

$$N(0)T_{t_I} = \sin^2\frac{\theta}{2}\cos\varphi\left[\left(-3.3 + 2.28\cos^2\frac{\theta}{2} - 5.82\cos^4\frac{\theta}{2}\right) - 0.74\sin^4\frac{\theta}{2}(5\cos^2\varphi - 3)\right] \quad (37)$$

$$N(0)T_{t_{II}} = -\left[4.78 + \theta^2\left(0.54\cos^2\frac{\varphi}{2} - 3.05\sin^2\frac{\varphi}{2}\right)\right] \quad (38)$$

$$N(0)T_{t_{III}} = \left[4.78 + \theta^2\left(0.54\sin^2\frac{\varphi}{2} - 3.05\cos^2\frac{\varphi}{2}\right)\right] \quad (39)$$

It is noted that in obtaining the above equations we truncate the sum in Eq. (15) for $k = 3$ at melting pressure. Of course, it should be mentioned that the s-p approximation with the Landau sum rule fulfilled is equivalent to truncate the $k$ sum in Eq. (15) at $k = 1$ [21]. Finally we obtain

$$\lambda_{2s_\downarrow} \simeq \lambda_{2s_\uparrow} \simeq 0.75 \quad (40)$$

$$B_{2s_\uparrow} \simeq \frac{\pi^5}{m^{*3}T^2}\frac{N(0)^2}{186.53} \quad , \quad B_{2s_\downarrow} \simeq \frac{\pi^5}{m^{*3}T^2}\frac{N(0)^2}{195.81} \quad (41)$$

Following the Sykes et al[25] procedure, the following result is obtained from Eq. (32)

$$\int_{-\infty}^{+\infty} dt \frac{dv^0}{dt} q_{2s_\sigma}(t) = \frac{-2B}{\pi^2(1-\lambda_{2s_\sigma})} c(\lambda_{2s_\sigma}) \quad (42)$$

where $c(\lambda)$ is [25]

$$c(\lambda_{2s_\sigma}) = \frac{(1-\lambda_{2s_\sigma})}{4}\sum_{n=0}^{\infty}\frac{(4n+3)}{(n+1)(2n+1)\{(n+1)(2n+1) - \lambda_{2s_\sigma}\}}$$

$$= \frac{\lambda_{2s_\sigma} - 1}{2\lambda_{2s_\sigma}}\left\{\gamma + \ell n 2 + \frac{1}{2}\psi_d(s_1) + \frac{1}{2}\psi_d(s_2)\right\} \quad (43)$$

and $\gamma = 0.577...$ is Euler's constant, $\psi_d$ is a digamma function and

$$s_1 \equiv \frac{3}{4} + \frac{1}{4}\sqrt{8\lambda_{2s_\sigma} + 1} \quad , \quad s_2 \equiv \frac{3}{4} - \frac{1}{4}\sqrt{8\lambda_{2s_\sigma} + 1} \quad (44)$$

and we have $c(\lambda_{2s_\uparrow}) \simeq c(\lambda_{2s_\downarrow}) \simeq 0.77$.

Now we will proceed to calculate the shear viscosity in the next section by using Eq. (42) and the calculated values of $B_\sigma$, $\lambda_{2s_\sigma}$ and $c(\lambda_{2s_\sigma})$.

## 4- Viscosity

The shear viscosity in general is a fourth-rank tensor which is defined by the relation

$$\pi_{lm} = -\sum_{ik} \eta_{lmik} \left( \frac{\partial u_i}{\partial r_k} + \frac{\partial u_k}{\partial r_i} - \frac{2}{3} \delta_{ij} \vec{\nabla} \cdot \vec{u} \right) \tag{45}$$

where $\pi_{lm}$, the momentum flux tensor, is

$$\pi_{lm} = \int d\tau_p p_l \frac{\partial E}{\partial p_m} \delta \nu_\sigma(p) \tag{46}$$

when Eq.(28) is substituted in Eq.(24) and then in Eq.(46) and compared with (45), we get

$$\eta_{lmik} = -\frac{4 p_F^5}{(2\pi)^3 m^*} \int d\Omega_p \, \hat{p}_l \hat{p}_m \hat{p}_i \hat{p}_k \left( \int dt \frac{\partial \nu^0}{\partial t} q_{2s_\uparrow}(t) + \int dt \frac{\partial n^0}{\partial t} q_{2s_\downarrow}(t) \right) \tag{47}$$

By using (42) in Eq. (47), we have

$$\eta_{lmik} = \frac{p_F^5}{\pi^5 m^*} \int d\Omega_p \, \hat{p}_l \hat{p}_m \hat{p}_i \hat{p}_k \left( \frac{B_\uparrow}{1-\lambda_{2s_\uparrow}} c(\lambda_{2s_\uparrow}) + \frac{B_\downarrow}{1-\lambda_{2s_\downarrow}} c(\lambda_{2s_\downarrow}) \right) \tag{48}$$

where we employ the fact that at low temperature all momenta lie on the Fermi surface.

After substituting $B_\sigma$, $\lambda_{2s_\sigma}$ and $c(\lambda_{2s_\sigma})$ and taking the angular integrations, we have

$$\eta_{xy_\downarrow} = \eta_{xz_\downarrow} = \frac{1}{3} \eta_{zz_\downarrow} \simeq \frac{N(0)^2 p_F^5}{m^{*4}} 13.17 \times 10^{-3} \frac{1}{T^2} \tag{49}$$

$$\eta_{xy_\uparrow} \simeq \frac{N(0)^2 p_F^5}{m^{*4}} 4.28 \times 10^{-4} \frac{T^4}{T_c^6} \tag{50}$$

$$\eta_{xz_\uparrow} \simeq \frac{N(0)^2 p_F^5}{m^{*4}} 8.16 \times 10^{-4} \frac{T^2}{T_c^4} \tag{51}$$

$$\eta_{zz_\uparrow} \simeq \frac{N(0)^2 p_F^5}{m^{*4}} 10.36 \times 10^{-4} \frac{1}{T_c^2} \tag{52}$$

By nothing that $\eta_{xy} \equiv \eta_{xy_\downarrow} + \eta_{xy_\uparrow}$, $\eta_{xz} \equiv \eta_{xz_\downarrow} + \eta_{xz_\uparrow}$ and $\eta_{zz} \equiv \eta_{zz_\downarrow} + \eta_{zz_\uparrow}$, finally we have

$$\eta_{xy} = \eta_{xz} = \frac{1}{3} \eta_{zz} \simeq \frac{N(0)^2 p_F^5}{m^{*4}} 13.17 \times 10^{-3} \frac{1}{T^2} \tag{53}$$

Now we proceed to express Eq. (53) in terms of the shear viscosity at $T_c$, $\eta(T_c)$. $\eta(T_c)$ is $(1/5)\rho(m^*/m)v_F^2\tau_0$ with $\tau_0$ being a characteristic relaxation time and given by $8\pi^4/m^{*3}T_c^2\langle W_N\rangle$ and density of the liquid, $\rho$, being $mp_F^3/3\pi^2$. $\eta(T_c)$ depends on $\langle W_N\rangle$, $m^*$, $v_F$, $T_c$ and $\tau_0$. In the following we calculate the constant prefactor by using different values of these quantities and different methods.

Case 1:

If we use the Pfitzner procedure, after some algebra we have

$$\langle W_N\rangle = \int \frac{d\Omega}{4\pi\cos\frac{\theta}{2}}\frac{2\pi}{8}\left(3|T_t|^2+|T_s|^2\right)\simeq\frac{130.56}{N(0)^2} \tag{54}$$

then we have

$$\eta_{xy}=\eta_{xz}=\frac{1}{3}\eta_{zz}\simeq 0.33\left(\frac{T}{T_c}\right)^{-2}\eta(T_c) \tag{55}$$

Case 2:

If we use s-p approximation for calculating singlet and triplet quasiparticle scattering amplitude and the numerical values of Landau parameters in [1] ([27]) for all step of calculating $\eta$, then we get the prefactor in Eq. (55) as equal to 0.15 (0.18).

Case 3:

If we use the values of $\eta(T_c)$ and $T_c^2$ from [19], $v_F$ from [28], and $m^*/m$ from [29], we get

$$\eta_{xz}\simeq 0.34\left(\frac{T}{T_c}\right)^{-2}\eta(T_c) \tag{56}$$

It is amusing that if we take $m^*/m$ from [28] the above prefactor will be 0.47.

Case 4:

If we use the values of $\tau_0 T_c^2$ and $v_F$ from [28] and $m^*/m$ from [29], we get

$$\eta_{xz}\simeq 0.44\left(\frac{T}{T_c}\right)^{-2}\eta(T_c) \tag{57}$$

It is amusing that if we take $m^*/m$ from [28] the above prefactor will be 0.47.

## 5. Conclusion

By supposing that Bogoliubov quasiparticles with spin up are formed in the superfluid component in $A_1$–phase, we calculate the transition probabilities for the cases where both the normal and Bogoliubov quasiparticles are presented in decay, coalescence and binary processes, beyond s-p approximation. We use the Pfitzner [21] procedure for obtaining singlet and triplet quasiparticle scattering amplitude. Then we calculate the components of shear viscosity tensor of the $A_1$ – phase of superfluid $^3He$ at low temperatures and high magnetic field at melting pressure.

Roobol et al [20] results on the shear viscosity of the $A_1$ – phase of superfluid $^3He$ in a static magnetic field up to 15 T at low temperatures indicate $\eta = 0.48\eta(T_c)(T/T_c)^{-2}$. They mention that in measuring the values of the viscosity, $\eta$, mainly the values of the component $\eta_{xz}$ with a few percent of the values of the component $\eta_{zz}$ will be contributed. If we take the contribution of $\eta_{zz}$ in $\eta$ between 5 and 10 percent, we obtain the prefactor between 0.44 and 0.40 respectively. This result is in fairly good agreement with our result which is obtained in case 4. In this case we benefit from the Pfitzner procedure plus the values of the quantities $\tau_0 T_c^2$ and $v_F$ from [28] and $m^*/m$ from [29]. In the following it would be suitable to present a quantitative comparison between results obtained from the Pfitzner procedure and s-p approximation. While the calculated prefactor for case 4 through s-p approximation is about 0.18, and considering the fact that the calculated figure has an error margin of a hundred percent and even more in comparison with the experimental data, the calculated prefactor for case 4 through Pfitzner procedure is 0.44 having a few percent margin of error. Hence, Pfitzner procedure gives better results than s-p approximation. It is noted that this is also the case for the results near $T_c$ [11] where Pfitzner procedure was implemented for the first time to calculate one of the transport coefficients of superfluid $^3He$. Moreover, implementation of this procedure in calculation of the transport times of normal fluid has previously proved to give better and more acceptable quantitative results [21]. On the whole one can say that Pfitzner procedure give more satisfactory results than s-p approximation.

We obtain $\eta_{xy\uparrow}$, $\eta_{xz\uparrow}$ and $\eta_{zz\uparrow}$ with dependence temperature $T^4$, $T^2$ and $T^0$ respectively (see Eqs.(50)-(52) ). It is seen that $\eta_\uparrow$ do not play a rule in the shear viscosity components compare to $\eta_\downarrow$. It is interesting to mention that interaction between superfluid and normal fluid in calculating the transition probabilities is distinct, whereas the contribution from the interaction between Bogoliubov quasiparticles at low temperatures is ignorable. Also it is mentioned that just binary processes are dominated in calculating the transition probabilities.

In summery we can say that the viscosity components at low temperatures in the $A_1-$phase of superfluid $^3He$ are proportional to $T^{-2}$ and the agreement between our results and the experiments is fairly good.

References:

[1] D.Vollhardt, P. Wölfle, The Superfluid Phases of Helium 3, Taylor and Francis Inc., London, 1990.

[2] P. W. Anderson, W. F. Brinkman, Phys. Rev. Lett. 30 (1973) 1108.

[3] G. Barton, M. A. Moore, J. Phys. C: Solid State phys. 8 (1975) 970.

[4] V. Ambegaokar, N. D. Mermin, Phys. Rev. Lett. 30 (1973) 81.

[5] V. S. Shumeiko, Zh. Eksp. Teor. Fiz. 63 (1972) 621 [English transl., Soviet Phys- JETP 36 (1973) 330].

[6] C. J. Pethick, H. Smith, P. Bhattacharyya, J. Low Temp. Phys. 23 (1976) 225.

[7] P. Bhattacharyya, C. J. Pethick, H. Smith, Phys. Rev. B 15 (1977) 3367.

[8] E. J. Pethick, H. Smith, P. Bhattacharyya , Phys. Rev. B 15 (1977) 3384.

[9] M. A. Shahzamanian, J. Phys. C: Solid State Phys. 21 (1988) 553.

[10] M. A. Shahzamanian, J. Low Temp. Phys. 21 (1975) 589.

[11] M. A. Shahzamanian, R. Afzali, J. Phys.: Condens. Mather 15 (2003) 367.

[12] C. N. Archie, T. A. Alvesalo, J. D. Reppy, R. C. Richardson, J. Low Temp. Phys. 42 (1981) 295.

[13] D. C. Carless, H. E. Hall, J. R. Hook, J. Low Temp. Phys. 50 (1983) 605.

[14] T. A. Alvesalo, Yu. D. Anufriyev, H. K. Collan, O. V. Lounasmaa, P. Wennerstrom, Phys. Rev. Lett. 30 (1973) 962.

[15] T. A. Alvesalo, H. K. Collan, M. T. Loponen, M. C. Veuro, Phys. Rev. Lett. 32 (1974) 981.

[16] T. A. Alvesalo, H. K. Collan, M. T. Loponen, O. L. Lounasmaa, M. C. Veuro , J. Low Temp. Phys. 19 (1975)1.

[17] R. T. Johnson, R. L. Kleinberg, R. A. Webb, J. C. Wheatley, J. Low Temp. Phys. 18 (1975) 501.

[18] L. P. Roobol, R. Jochemsen, C. M. C. M. Van Woerkens, T. Hata, S. A. J. Wiegers, G.

Frossati, Physica B 165 and 166 (1990) 639.

[19] L. P. Roobol, W. Ockers, P. Remeijer, S. C. Steel, R. Jochemsen, G. Frossati, Physica B 194-196 (1994) 773.

[20] L. P. Roobol, P. Remeijer, S. C. Steel, R. Jochemsen, V. S. Shumeiko, G. Frossati, Phys. Rev. Lett.79 (1997) 685.

[21] M. Pfitzner, J. Low Temp. Phys.61 (1985) 141.

[22] S. Takagi, J. Low Temp. Phys.18 (1975) 309.

[23] G. Baym, C. Pethick, Landau Fermi-liquid Theory, John Wiley & Sons Inc., U.S.A, 1991.

[24] M. A. Shahzamanian, J. Phys.: Condens. Matter 1 (1989) 1965.

[25] J. Sykes, G. A. Brooker, Ann. Phys. 56 (1970) 1.

[26] D. D. Osheroff , P. W. Anderson, Phys. Rev. Lett. 33 (1974) 686.

[27] D. Einzel, P. Wölfle, J. Low Temp. Phys. 32 (1978) 19.

[28] J. Wheatley, Rev. Mod. Phys. 47 (1975) 415.

[29] D. S. Greywall, Phys. Rev. B 33 (1986) 7520.